\documentclass[preprint,12pt]{elsarticle}

\usepackage{amssymb}
\usepackage{amsmath}

\usepackage[utf8]{inputenc}

\usepackage{hyperref}
\usepackage{soul}
\usepackage{xcolor}
\usepackage{xspace}
\usepackage{url}
\usepackage{graphicx}
\usepackage{listings,multicol}
\usepackage[caption=false]{subfig}  
\usepackage[export]{adjustbox}
\usepackage{amsmath}
\usepackage{textcomp}
\usepackage{cleveref}
\usepackage{algorithm}
\usepackage[noend]{algpseudocode}
\usepackage{diagbox}
\usepackage{fixfoot}
\usepackage{multirow}
\usepackage{booktabs}
\usepackage{colortbl}
\usepackage{arydshln}
\usepackage{comment}

\definecolor{gray}{gray}{0.9}

\usepackage{balance}
\newcounter{observation}
\newcommand{\observation}[1]{\refstepcounter{observation}
	\begin{center}
		\framebox{
			\begin{minipage}{0.93\columnwidth}
				{} \textit{#1}
			\end{minipage}
		}
	\end{center}
}

\definecolor{APA_stats}{RGB}{100, 100, 120}
\newcommand{\APAstats}[2]{\textcolor{APA_stats}{(M=#1, SD=#2)}}

\newcommand{\APAr}[2]{\textcolor{APA_stats}{r=#1, p $\leq$ #2}}

\def\Nstudents{11\xspace}

\def\Nstudents{11\xspace}
\def\Nprof{9\xspace}
\def\Ncells{2,655\xspace}
\def\Ntasks{29\xspace}
\def\Nevents{14,641\xspace}
\def\Nexecutions{9,207\xspace}
\def\Ncreations{1,930\xspace}
\def\Ndeletions{730\xspace}

\journal{Journal of Systems and Software}

\begin{document}

\begin{frontmatter}

\title{Observing Fine-Grained Changes in Jupyter Notebooks During Development Time}

\author[JB1]{Sergey Titov\corref{mycorrespondingauthor}}
\cortext[mycorrespondingauthor]{Corresponding author}
\ead{sergey.titov@jetbrains.com}

\author[JB2]{Konstantin Grotov}
\ead{konstantin.grotov@jetbrains.com}

\author[UZH]{Cristina Sarasua}
\ead{sarasua@ifi.uzh.ch}

\author[JB3]{Yaroslav Golubev}
\ead{yaroslav.golubev@jetbrains.com}

\author[UZH]{Dhivyabharathi Ramasamy}
\ead{ramasamy@ifi.uzh.ch}

\author[UZH]{Alberto Bacchelli}
\ead{bacchelli@ifi.uzh.ch}

\author[UZH]{Abraham Bernstein}
\ead{bernstein@ifi.uzh.ch}

\author[JB4]{Timofey Bryksin}
\ead{timofey.bryksin@jetbrains.com}

\address[JB1]{JetBrains Research, Amsterdam, The Netherlands}
\address[JB2]{JetBrains Research, Munich, Germany}
\address[JB3]{JetBrains Research, Belgrade, Serbia}
\address[JB4]{JetBrains Research, Limassol, Cyprus}
\address[UZH]{University of Zurich, Zurich, Switzerland}

\begin{abstract}

In software engineering research, the analysis of fine-grained logs led to significant innovations in areas such as refactoring, security, and code completion. However, even though computational notebooks are a staple of data science and an important tool in machine learning, few similar studies have been conducted in this area.

To help bridge this research gap, this paper makes three scientific contributions. (1) We introduce a \emph{toolset} for collecting code changes in Jupyter notebooks during development time. (2) We use it to collect more than 100 hours of work related to a data analysis task and a machine learning task (carried out by 20 developers with different levels of expertise), resulting in a \emph{dataset} containing 2,655 cells and 9,207 cell executions. (3) Finally, we use this dataset to \textit{investigate} the dynamic nature of the notebook development process and the changes that take place in the notebooks.

In our analysis of the collected data, we classified the changes made to the cells between executions and found that a significant number of these changes constituted code iteration modifications. We report a number of other insights and propose detailed future research directions on the novel data.

\end{abstract}


\end{frontmatter}

\section{Introduction}

The analysis of fine-grained software development logs provides information about the development process at a higher level of detail compared to the analysis of version control system (VCS) snapshots~\cite{negara2012dangerous}. In the last several decades, software engineering research has employed development logs to study, for example, program security~\cite{ko1997execution}, performance~\cite{syer2013leveraging}, and code completion~\cite{bibaev2022all}. In particular, \textit{fine-grained execution logs} that save the information about various activities during the code writing process~\cite{lyulina2021tasktracker} have opened up the possibility of studying developer behavior, \textit{e.g.}, how the integrated development environment (IDE) affects developers with ADHD~\cite{kasatskii2023effect}. 

When it comes to computational notebooks, however, there are almost no studies or datasets on their fine-grained evolution or execution during development time.
For this reason, 
most of the studies analyze computational notebooks using examples downloaded from public version control systems. For example, Pimentel et al.~\cite{pimentel2021understanding} show that a significant portion of notebooks are not reproducible due to incorrect cell ordering. Grotov et al.~\cite{grotov2022large} compare notebooks to Python scripts and find that notebooks generally have lower code quality.
Yet, having fine-grained data on notebook development and evolution is critical, because notebooks represent an alternative paradigm of literate programming~\cite{knuth1984literate} that possesses unique execution characteristics. 
This was brought forward by the analysis of fine-grained VCS logs in the work of Raghunandan et al.~\cite{raghunandan2023code}, where the authors examine multiple historical versions of the same notebook from version control systems and demonstrate that notebooks can change drastically between versions.

One study based on truly fine-grained logs is the work of Chen et al.~\cite{chen2025towards}, which explores execution logs in Jupyter Notebooks. The authors analyze notebook executions to identify the types of errors made by data science practitioners. The fine-grained resolution of these logs allowed them to examine the specific strategies users employ to fix errors and to suggest ways to improve automated program repair.

To go further and broaden this research, we propose three contributions.
As our first contribution, we devised the tooling necessary to obtain fine-grained log data. The main tool is a plugin for the Web-based Jupyter Notebook IDE that tracks and logs executions of the opened Jupyter notebook upon every action. Additionally, we developed a server-side application that receives and stores this data, as well as a collection of post-processing scripts for parsing, analyzing, and visualizing the collected data. The tooling is designed to be convenient to use, adaptive to running live experiments, and easy to extend, leading to further data collection and studies in this area. It is available online~\cite{artifacts}.

As our second contribution, using the developed tools, we collected the dataset of fine-grained execution logs for Jupyter notebooks --- Jupyter Notebooks Executions (\textbf{JuNE}). To this end, we organized a series of day-long experiments where we asked participants to solve at least one of two development tasks --- one related to data analysis and one related to machine learning. Overall, \Nstudents students (enrolled in Master's and Bachelor's studies that include Data Science-related topics) and \Nprof industry professionals (with a minimum of one year of experience in a Data Science-related role) participated and sent us their data.
The resulting dataset contains the information about a total of \Ncells developed cells and \Nexecutions cell executions, and includes timestamps, the content of the cells on every execution, the labeling of the data science steps in the cell using an annotation model, the type of the implemented action (\textit{e.g.}, cell creation, cell deletion, or cell execution), and the cell execution order. We release this dataset to enable further studies taking advantage of these fine-grained execution logs~\cite{artifacts}.

As our third contribution, having obtained the dataset, we carried out a fine-grained analysis of the development process in Jupyter notebooks. We focus on the code changes made to cells during the iterative process of editing and execution. By analyzing this data, we investigate how the creation and evolution process in the notebooks unfolds---what notebook users do while working on cells and how we can support them more effectively. 

In order to analyze changes, we transformed the dataset to represent notebook development as a series of transitions from one cell to another. 
We identified two types of transitions: \textit{inter-transitions}, where the developer moves from one cell to another, and \textit{self-transitions}, where the developer re-executes the same cell.  We found that 39\% of transitions are self-transitions, while the remaining 61\% are inter-transitions. Self-transitions represent the continuous development of a cell and will be the primary focus of our empirical analysis, as we are interested in code changes.

Additionally, we introduced two new annotations for the cells. First, we annotated each self-transition with one of 11 possible \textit{purposes} (\textit{e.g.}, ``fix'', ``clean code'', ``explore variable'', etc.). We conducted a manual open coding on 400 examples and then used GPT-4o to annotate the remaining transitions. Second, we annotated each cell with its corresponding \textit{data science step}, based on the classification introduced by Ramasamy et al.~\cite{ramasamy2023workflow} (\textit{e.g.}, ``evaluation'' or ``load data''). To annotate the data science steps, we developed our own model tailored to our data format, which performs slightly better than the original.

Through our analysis we observe that the changes in notebooks are on average relatively small (with only about 13\% of a cell being modified) and that this size decreases with consecutive re-executions of the same cell. We found that most changes revolve around the process of code iteration, with 72.2\% of the change purposes in self-transitions aimed at fixing, debugging, editing, cleaning code, improving readability, etc., while exploration is a relatively rare goal, accounting for only 14.1\% of transitions. Lastly, we found evidence that data science steps are highly stable during self-transitions---the type of a cell rarely changes during cell development, but changes more frequently in the case of inter-transitions.

The results lead us to hypothesize that the interactive nature of notebooks is leveraged for debugging rather than for exploration with rich outputs. This highlights the need for more convenient debugging tools and IDE-like inspection features to check the code prior to execution.

In summary, this work makes the following main contributions:

\begin{itemize}
    \item \textbf{Tooling} for collecting fine-grained execution logs in Jupyter notebooks, consisting of a plugin to capture user activity, a server to receive and store the data, and a collection of post-processing scripts to parse and analyze it. We also provide a tool to replay notebook executions, effectively allowing to explore the notebook’s state at various points. Our data collection effort also ensured that the tooling can handle real-world, extensive usage. 
    \item \textbf{Dataset} of fine-grained execution logs that we collected from \Nstudents students and \Nprof industry professionals. In total, the dataset contains the data about \Ncells developed cells and \Nexecutions cell executions---for a total of more than 100 hours of data science and machine learning related work---together with detailed information about their evolution and content. 
    \item \textbf{Empirical findings} about the changes that occur in notebooks during their development, which reveal important insights. We annotated the collected data based on the purpose of the changes in code transitions and the data science step they refer to. Our analysis shows that the development process in notebooks is highly non-linear and lacks proper support for development tools. We suggest that further analysis of this data could provide valuable insights for improving notebook tooling.
\end{itemize}

The remainder of the paper is organized as follows. In Section~\ref{sec:background}, we describe the existing work studying Jupyter notebooks and collecting fine-grained software logs. In Section~\ref{sec:tooling}, we present the tooling we developed for collecting fine-grained logs in Jupyter notebooks. We detail our data collection process in Section~\ref{sec:experiment} and then present the obtained data in Section~\ref{sec:dataset}. In Section~\ref{sec:features}, we describe the methodology of our analysis, and present the findings in Section~\ref{sec:findings}. Then, we discuss the implications of our study and the threats to its validity in Section~\ref{sec:discussion}. Finally, we present future research directions in Section~\ref{sec:fututre} and conclude in Section~\ref{sec:conclusion}.
\section{Background and Related Work}
\label{sec:background}
\subsection{Code Evolution in Software Engineering}

More than ten years ago Negara et al.~\cite{negara2012dangerous} raised questions about the reliability of VCS data for investigating software evolution and what it lacks. Using an Eclipse plugin, they collected fine-grained development logs and compared them with VCS data from the same project. They found that 37\% of changes occur during development and never reach VCS. Additionally, they discovered insights in development logs that are difficult to glean from VCS data alone. For example, they found that 24\% of the changes eventually committed to VCS had not been tested prior to their inclusion.

The analysis of development logs has led to multiple innovations in the current generation of IDEs. For example, in another paper~\cite{negara2013comparative}, the same authors provide a comprehensive analysis of refactoring techniques employed by developers. They demonstrated that half of all refactorings are done manually, even though many could be automated by the IDE. For instance, their study highlights that the frequency of automated \textit{Extract Method} refactoring does not correlate with the size of the code, suggesting usability issues. Developers tend to make manual changes, even when they involve rewriting larger pieces of code.

Another significant work in the field is the study of Yoon et al.~\cite{yoon2014longitudinal}, in which the authors provided evidence of a significant number of backtracking actions during the development process and a lack of tools to support them. The authors collected log data about code editing in Eclipse and characterized the different types of backtracking, as well as the frequency and size of the backtracking. They also proposed a tool to selectively undo the desired previous edits without undoing certain intermediate changes.

There are other studies that leverage development logs for code completion~\cite{bibaev2022all}, various types of refactoring~\cite{negara2014mining}, as well as test design and generation~\cite{hilton2016tddviz}.
Due to the changes in the availability of detailed development logs, the number of studies conducted in this area has recently declined. With increasing technical complexity of the systems, as well as legal complexity of the entire field, collecting such data becomes more problematic~\cite{van2019possibilities}.

At the same time, collecting and analyzing fine-grained development logs in the context of Jupyter Notebooks is crucial for advancing our understanding of how notebooks are authored, used, and evolved over time. Unlike traditional software development environments, notebooks offer a unique blend of narrative, code, and execution semantics, which likely leads to distinct development and maintenance practices. For instance, it remains unclear whether and how refactoring strategies differ in notebooks. As observed by Titov et al.~\cite{titov2022resplit}, notebook cells can be conceptualized as proto-functions. Fine-grained logs could thus reveal notebook-specific refactoring patterns---such as the extraction or reorganization of cells---that are not observable through coarse-grained data alone.

\subsection{Coding in Jupyter Notebooks}

Studies of computational notebooks are an established subfield in software engineering research~\cite{wang2020better}, with important empirical insights. Research showed that notebooks have a lower rate of reproducibility~\cite{pimentel2019large, pimentel2021understanding}, significantly differ in terms of code structure from Python scripts~\cite{grotov2022large}, and frequently contain code clones~\cite{kallen2020jupyter}. These results show that the analysis of code in notebooks is challenging, which affects the performance of tools for assisting developers, such as code completion or code linting.

Recently, Ramasamy et al.~\cite{ramasamy2023workflow} analyzed the developers' workflow in 470 notebooks and showed the probability of transitions between different data science steps in subsequent cells of the final snapshot of a notebook (\textit{e.g.}, data pre-processing, modeling, evaluation, etc.). The results revealed that 53.25\% of the code cells in the analyzed data science workflows focused on the tasks of data pre-processing and data exploration.
In our paper, we used the classification of data science steps from this work because this taxonomy is the most complete. Other taxonomies, \textit{e.g.}, from the work of Wang et al.~\cite{b1_Wang_2019}, do not distinguish between model training and model evaluation, while loading the data is not captured at all. 

There were also attempts to look at the notebook's evolution over time. Raghunandan et al.~\cite{raghunandan2023code} looked at how a set of 2,574 notebooks changed over different commits on GitHub. More specifically, they analyzed the extent to which the notebooks change or maintain their purpose (exploratory or explanatory). Their study revealed that most of the notebooks in their dataset (60.1\%) keep their explanatory nature over the commits, 22.6\% keep their exploratory purpose, while 15.1\% of the notebooks went from being exploratory to being explanatory, and only 2.1\% transitioned from being explanatory to being exploratory. 

In a recent work, Chen et al.~\cite{chen2025towards} analyzed fine-grained Jupyter notebook traces. The authors collected fine-grained execution logs to develop a taxonomy of the errors that occur during data science work and to examine how users fix them.
Their analysis shows that most errors during notebook development occur in the data exploration and preprocessing phases, and are usually fixed by changing parameters or correcting syntax errors. They argue that these findings could significantly improve automated program repair for Jupyter notebook use cases, since the range of errors is relatively small.

We believe that this field needs to be extended further to move away from coarse-grained commit-centered dynamics and support better developer tooling. Our work focuses on wider patterns, not just errors, presents mature tooling suitable for large-scale experiments and replaying collected notebooks, a novel dataset, and more empirical findings about the nature of work in Jupyter notebooks.
\section{Tooling}
\label{sec:tooling}

\begin{figure}
    \centering
    \includegraphics[width=\textwidth]{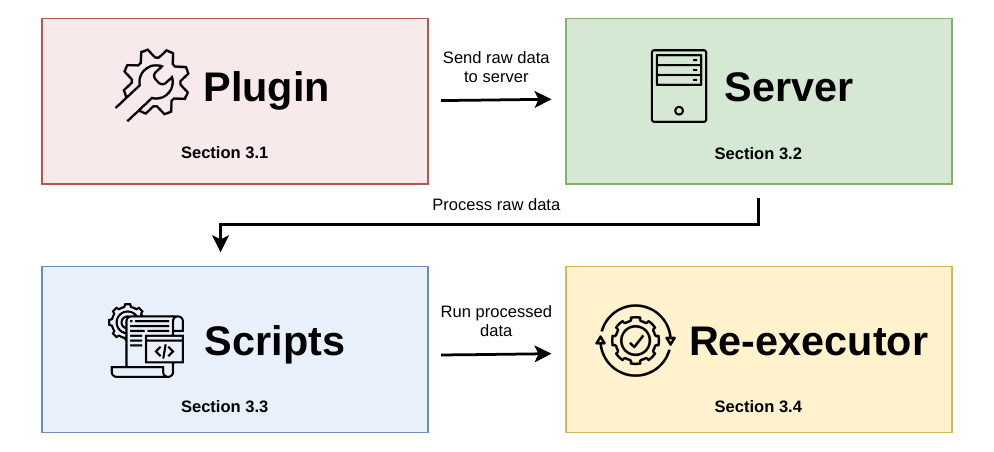}
    \caption{The pipeline of the developed tooling.}
    \label{fig:tooling}
\end{figure}

To collect fine-grained execution logs of Jupyter notebooks and to conduct our empirical study, we developed the necessary tooling. It consists of four main parts, visualized in Figure~\ref{fig:tooling}:

\begin{itemize}
    \item A \textbf{plugin} to collect the necessary data.
    \item A \textbf{server} to receive and store the data.
    \item A collection of \textbf{post-processing scripts} to process the data and conduct the analysis.
    \item A \textbf{tool to re-execute notebooks} from obtained logs, given the correct environment.
\end{itemize}

\subsection{Activity Plugin}
\label{sec:plugin}

To collect user activity, we developed a JavaScript plugin as an extension for the Jupyter Notebook Web application. The plugin tracks the initial launch of a Python notebook, as well as its interruption and restarting. Additionally, the extension tracks certain events that occur during the development process in the notebook, including \emph{creating a new cell}, \emph{deleting a cell}, \emph{executing a cell}, \emph{rendering a Markdown cell}, \emph{finishing executing}, \emph{changing a cell type} (from code to Markdown or vice versa), and \emph{errors} (when the executed cell finished by raising an error). For each action, we save all the possible information about the cell, the notebook, and the user. For example, when a cell is executed, the plugin collects the index of the cell, its ordinal number in the notebook, and the cell's source code. Also, for all types of events, we save information about the session, the notebook kernel, the notebook name, and the timestamp.
Upon collection, the data is stored locally and, if necessary, sent to a remote server, the address of which can be set up in the settings. 

During the development of the plugin, our main goal was to minimize user interaction to ensure that it does not influence the coding process.
To begin using the plugin, the users first need to install it as a Python package. Once installed, they activate it as a Jupyter extension. Before starting the plugin, the users are prompted to consent to send data to a remote server and provide the server's address.
Once these initial steps are completed, the plugin transparently runs in the background, collecting data across all Jupyter notebooks that are accessible to the given Jupyter Notebook application instance, without interrupting the user. This supports authentic studies that can provide realistic insights into developers' workflows.

\subsection{Server}
To handle the data generated and sent by the plugin, we developed a remote Python server using Flask~\cite{grinberg2018flask}. The server receives the data using a \texttt{get} request and saves it to a local SQLite database~\cite{owens2010sqlite}. When the experiment is finished, the server is shut down. The presence of a dedicated server helps run large-scale experiments, without necessarily relying on manually sending and saving files.

\subsection{Post-processing Scripts}

The data collected from the plugin consists of a collection of raw events from all participating users. Due to the internal implementation of the Jupyter Notebook Web application, one can find inconsistencies in the raw data. For example, Jupyter kernel (Notebook version 6.4.12) sends the \textit{rendered} status for a new Markdown cell before the creation of the corresponding cell. To address possible issues, we developed a set of Python scripts that perform post-processing on the data using a number of heuristics. This package allows for more convenient and efficient analysis of the collected data by transforming it from raw JSON files into two table formats: a log table and a set of notebook snapshots. The log table contains every action for each user of the plugin in chronological order. Snapshot tables serve as a set of snapshots, preserving the entire sequence of notebook cells after each action recorded in the log table. These transformations enable the use of various Python libraries. For example, they enable the analysis of the entire log of the notebook as a time series of actions, or the analysis of each snapshot of the notebooks to track the evolution of the code.

\subsection{Browsing the Notebook Versions}
\label{sec:historytool}

A key benefit of the data collected with our tools is that it allows researchers to reproduce the developers' workflow in the notebook. To do that, we developed a tool for setting up the necessary environment and then reproducing the notebook to browse its particular version. After the user selects a specific collected notebook, a new Jupyter notebook is created within the separate Docker container. Each development log corresponds to a notebook workflow and is translated into actions within this environment, effectively reproducing the developer’s actions. This enables the analysis of not only the code but also the state of the environment at each step of the collected data. For example, this could help train a machine learning model that relies on runtime information for code completion. 
\section{Data Collection}
\label{sec:experiment}

The data collection procedure for our empirical study was designed around two variables that could influence the notebook development process: 
(1) the \textit{type} of the solved task and 
(2) the level of participants' \textit{expertise}. 
We designed two distinct tasks that demanded a wide spectrum of data science skills and, to analyze how the level of expertise impacts the development process, specifically targeted two groups of people: computer science students and industry data science professionals.

\subsection{Designing Tasks}

We developed two specific tasks for the experiment, designed to emulate the main activities found in Jupyter notebooks: a data analysis task (DA) and a machine learning task (ML). Our principal design goal was to create tasks that replicated the development and usage of an authentic notebook, both in terms of content and duration. We estimated each task to take around four hours, totaling eight hours---the approximate working day for many data scientists. To estimate completion time, the first two authors of this paper tested how long each task would take and adjusted them accordingly.

To encourage the participants to create good quality solutions, we informed them that all solutions would be scored, and we made available a public leaderboard for each task for the duration of the experiment. The full text of the tasks and the scoring criteria can be found in the supplementary materials~\cite{artifacts}. Let us now briefly describe them.

\subsubsection{Data Analysis Task}

For the DA task, participants were provided with a dataset of synthetic logs created by the authors, containing logs of user actions on some imaginary social network where users could post, like, or follow someone. The task consisted of three sub-tasks: (1) data engineering, (2) metrics evaluation, and (3) data visualization.

The first sub-task required participants to parse the input dataset into a designated table format and perform data cleaning. We injected the dataset with structural and syntactic errors in the log strings to test the participants' ability to identify and fix these errors. The second sub-task involved completing a set of statistical tasks, such as calculating the mean number of actions per user. Participants were given a total of five predefined statistical tasks to solve, along with an additional task where they had to devise their own metric of user activity based on the provided data. The third sub-task concerned the visualization of selected metrics. We instructed participants to create several distinct plots based on the provided data, including line plots, bar charts, and heatmaps. Additionally, we asked participants to create a visualization of any measurement they had computed that had not been visualized so far. 

Overall, participants were required to complete typical data analyst tasks, involving actions such as data cleaning, exploration, aggregation, and, ultimately, visualization. 

\subsubsection{Machine Learning Task} 

For the ML task, we created a Kaggle-like~\cite{bojer2021kaggle} competition. We used the already validated dataset and task instructions from the work of Ramasamy et al.~\cite{ramasamy2023workflow}. The dataset consists of 9,678 Jupyter notebook cells, categorized into one of ten possible classes by a group of five experts. The goal of this classification task is to automatically classify the type of data science steps present in Jupyter notebook cells. The dataset also included pre-computed features, such as the number of lines of code in a cell and the number of unique variables in each cell. The study participants were given access to both the train and test subsets of the dataset from the paper, and their final solutions were evaluated on a separate subset. Following the original procedure from the paper, the evaluation of the classification model was implemented using the weighted F1-score. 

The ML task included data science steps complementary to the steps present in the DA task: data exploration, data pre-processing, modeling, evaluation, and result visualization.

\subsection{Choosing Participants}

To recruit participants, we issued two calls for participation: one targeting students at two universities and one addressing employees at two companies. Interested individuals were asked to complete a short questionnaire about their Python experience and the frequency with which they use Jupyter notebooks. This helped us filter out participants with less than one year of Python experience or no Jupyter notebook experience. The final sample consisted of 20 people: 11 students and 9 industry professionals.

\subsection{Executing the Experiment}

Since the experiment took place in four different locations (two universities and two companies), we gathered all eligible participants into groups of two to nine people. Each person in the group was asked to solve at least one of the tasks, and participants implemented their notebooks individually. We provided each group with a repository containing task descriptions and the necessary data. Upon presenting the repository, we activated a nine-hour timer, which allowed approximately eight hours for task completion and an additional hour for lunch. After nine hours, we ceased logging and evaluated each solution based on the most recently logged notebook. Following our evaluation, we shared the leaderboard results with each group, as well as the aggregated leaderboard results of all previous groups.

\section{Dataset}
\label{sec:dataset}

The resulting dataset of Jupyter Notebooks Executions (\textbf{JuNE}) contains more than 100 hours worth of execution logs from 11 students and 9 professionals. Students and professionals differ not only in their formal status but also in their Python experience: \APAstats{56.3}{31.9} months for professionals and \APAstats{33.3}{11.5} months for students. 80\% of participants reported using Jupyter notebooks more than three times a week.

During the experiment, the participants solved a total of \Ntasks tasks, resulting in \Ntasks notebooks. This data included 16 solutions of the DA task and 13 solutions of the ML task. Overall, the data includes \Nevents user events, including \Nexecutions cell executions, \Ncreations cell creations, and \Ndeletions cell deletions. 

For each of the 29 notebooks, the dataset contains all the notebook's actions in chronological order, together with cell contents and all the meta-information. 
\section{Empirical Analysis Methodology} 
\label{sec:features}

Since our dataset offers a unique opportunity to examine the development process within notebooks, we decided to focus on this aspect by analyzing code changes. We seek new insights into how people use notebooks in real-world settings and to inform tool developers on how to support the development process. Our exploration of the code changes is structured around the following research questions:

\begin{itemize}
    \item \textbf{RQ1:} \textbf{How do code changes take place during cell development in Jupyter Notebooks?} Mining the execution traces of notebooks for changes provides a solid analysis of code evolution. More importantly, once collected, this data can offer far deeper insight into the Jupyter development process and notebooks as an environment than any stable Git snapshots.
    \item \textbf{RQ2:} \textbf{What are the purposes of these code changes?} To deepen our understanding of the discovered changes, we aim to develop a taxonomy based on our unique data. This will help refine the interpretation of the acquired data and provide a clearer picture of how notebooks evolve, as well as highlight the main reasons users interact with code in notebooks. 
    \item \textbf{RQ3:}\textbf{ How are the data science workflow steps represented during notebook development?} Lastly, we want to examine how the changes align with the data science development process and how the dynamic nature of our data reflects it. 
\end{itemize}

\begin{figure}
    \centering
    \includegraphics[width=\textwidth]{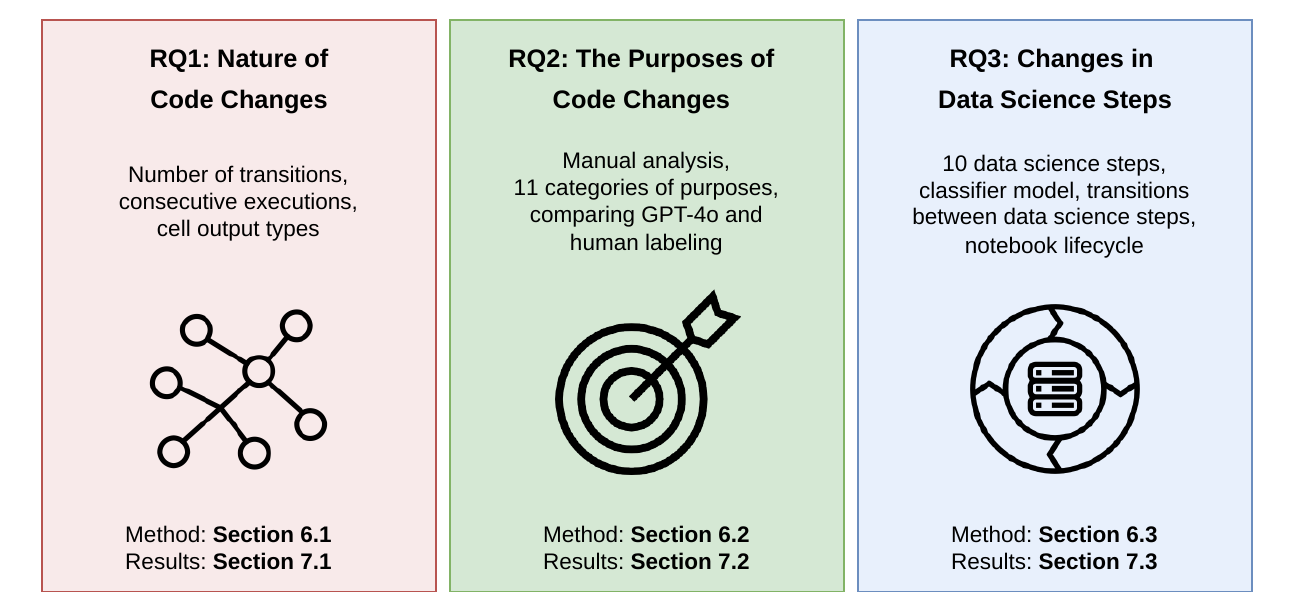}
    \caption{The overview of our empirical analysis.}
    \label{fig:method}
\end{figure}

The overview of the entire analysis can be found in Figure~\ref{fig:method}. Before conducting the analysis, we preprocessed the data to focus on the code changes. First, we filtered the dataset to retain only execution events, representing granular but logically finished steps. Then, we transformed the dataset to represent notebook development as a series of transitions from one cell to another. From the development log, we reconstructed the data as follows: for each execution, we recorded information about the state of the executed cell, including the time of the previous execution, the source code, and the saved outputs. We also stored the state of the next cell to be executed.
From this data, we identified two types of transitions: \textit{inter-transitions}, where the developer moves from one cell to another, and \textit{self-transitions}, where the developer re-executes the same cell.

To clarify the notion of transition, we consider the evolution graphs of the resulting notebooks. Figure~\ref{fig:example} shows an example of an evolution graph, where the nodes represent cells and edges represent transitions to the next executed cells. We can observe both transition types: self-transitions are highlighted in blue boxes, while inter-transitions are highlighted in red boxes.
Self-transitions represent the continuous development/evolution of a cell and are the primary focus of our empirical analysis, as we are interested in code evolution. In contrast, inter-transitions occur between different cells: they may represent the continuous development of a single idea, but they could also stand for two unrelated pieces of code. As it is not possible to automatically understand the reason why two cells are separated, we adopt a conservative approach and mostly focus on the self-transitions.

\begin{figure*}
    \centering
    \includegraphics[width=\textwidth]{figure-3-grapth-student.pdf}
    \caption{Example of structure of the evolution graph. Red boxes highlight examples of inter-transition and blue boxes highlight examples of self-transitions (loops).}
    \label{fig:example}
\end{figure*}

\subsection{RQ1: Nature of Code Changes}

To describe how source code evolves during notebook development, we calculate and present several metrics. 
First, we provide the overall number of inter-transitions and self-transitions. The distribution of the number of subsequent self-transitions (\textit{i.e.}, the lengths of re-execution chains) helps us determine how many times, on average, a cell is re-executed in a row, providing insight into the iterative nature of working in a notebook cell.
We also examine the distribution of edit distance between two consecutive executions, measured by the normalized number of changed symbols. This metric allows us to understand the average size of a change, revealing whether users tend to rewrite the cell entirely or make minor tweaks between executions.
Lastly, we calculate the distribution of cell output types (\textit{e.g.}, text, rich data, errors, etc.) before the self-transition. This analysis can offer insights into the reasons prompting the re-execution of a cell.

\subsection{RQ2: The Purposes of Code Changes}

To identify the purpose of the code changes, and thus the users' intentions when changing the code, we \emph{qualitatively} analyzed the code before and after self-transitions. The reason why in this part we focus exclusively on self-transitions is that inter-transitions consist of a developer executing two different cells (with two different pieces of code). Therefore, inter-transitions do not provide relevant information to analyze the evolution of the code. 

Due to the large scale of our dataset ($3,573$ self-transitions), we used a combination of manual open coding and automatic labeling using GPT-4o. We manually annotated a sample of $400$ self-transitions to define a set of labels that we could provide to GPT-4o in the prompts. Specifically, two authors of the paper manually reviewed $200$ transitions pertaining to the data analysis task and $200$ pertaining to the machine learning task. These transitions were randomly sampled following a stratified sampling based on the users, since some users generated more cells than others. Both annotators labeled each of these transitions following an open coding methodology, analyzing the source code of the cell before re-execution and the source code of the cell after re-execution. Subsequently, the two annotators discussed the resulting set of distinct labels (\textit{i.e.}, their labeling schemes) and identified mappings between them. While one of the annotators had a more fine-grained set than the other (\textit{e.g.}, one annotator defined the label ``modification'', while the other annotator had differentiated between ``correct to match what they aim for'', ``edit to a new variable version'', and	``test different input data''), we were able to map the labels with 1:1 and 1:N mappings and agree on the necessary specificity. 
As a result, we identified 11 labels shown in Table~\ref{tab:change-purpose-labels}.  

\begin{table}[]
    \small
    \centering
      
    \begin{tabular}{p{0.20\textwidth}p{0.80\textwidth}}
   
     \toprule
     \multicolumn{1}{c}{\textbf{Name}} & 
        \multicolumn{1}{c}{\textbf{Explanation}}\\
      \midrule

\rowcolor{gray}\texttt{no change} & The two pieces of code in the transition are identical. There is no change. \\
\rowcolor{white}\texttt{explore variable} & Code is added or edited to (further) explore the content or shape of a variable without changing the state of the code. In other words, there is no variable assignment involved in or affected by the code change. Operations may be applied to the content of a variable, but then the result is only displayed or printed, not assigned to a variable.  \\
\rowcolor{gray}\texttt{fix} & Code is edited or extended to fix a syntactic error. \\
\rowcolor{white}\texttt{debug} & Code is changed and/or added to 1) identify the source of a problem that leads to functional incorrectness or 2) to find a solution to the problem, or both 1) and 2). \\
\rowcolor{gray}\texttt{edit code} & Code is changed to adjust certain things and possibly obtain better results. This kind of change changes the state of code. In other words, there is a variable assignment involved in or affected by the code change. Please note that editing code to explore a variable in other ways should not be classified as `edit code', but instead it should be classified as a case of `explore variable'. \\
\rowcolor{white}\texttt{clean code} & Code was deleted.  \\
\rowcolor{gray}\texttt{visualize data} & A data visualization (e.g., a plot) is created or edited. \\
\rowcolor{white}\texttt{extend code} & Code that is extended with new partial or complete pieces of code to drive the development further. This kind of change changes the state of code. Please note that adding code to explore a variable further should not be classified as `extend code' but instead it should be classified as a case of `explore variable'. Analogously, adding code to create or refine a visualization should not be classified as `extend code' but instead it should be classified as a case of `visualize data'. \\
\rowcolor{gray}\texttt{improve readability} & Code was changed to improve the formatting of the code and make it easier for developers to read and understand it. \\
\rowcolor{white}\texttt{comment} & Comment code.  \\
\rowcolor{gray}\texttt{uncomment} & Uncomment code that was previously commented.  \\

    \bottomrule
    \end{tabular}
    \caption{Resulting classification for the purposes of the cell change after expert annotations. The selected labels were obtained by manually annotating 200 random transitions from each task (400 total), followed by a discussion until full agreement on the labels was reached.}
    \label{tab:change-purpose-labels}
\end{table}

As a second step, we programmatically labeled the self-transitions using GPT-4o. We provided GPT-4o with a system prompt that instructed the LLM to classify code transitions based on the purpose of the changes. Additionally, the system prompt provided our resulting labeling scheme and instructed that if GPT-4o found a case with a label that was not present in our scheme, GPT-4o should use its own label consistently across transitions. The user prompt focused on asking GPT-4o to label one code transition at a time. Both prompts together had a length smaller than the maximum context (\textit{i.e.}, 128k tokens), and, since they mostly duplicate the description of the classes provided in Table~\ref{tab:change-purpose-labels}, they can be found in supplementary materials~\cite{artifacts}. Two new labels were generated by GPT-4o (``unknown'' and ``remove debugging code''), but they were only present in one case each. 

Finally, as a third step, we revised the labels from the open coding together with GPT-4o's labels, to define the ground truth labels for the sample of 400 transitions. Comparing the three annotations simultaneously was useful, as in some cases, GPT-4o had provided the correct answer, while the human annotators had not---probably due to human factors, such as tiredness and lack of consistency.

\subsection{RQ3: Changes in Data Science Steps}
\label{sec:method:workflow}

\begin{table}[]
    \small
    \centering
    \begin{tabular}{p{0.3\textwidth}p{0.53\textwidth}p{0.10\textwidth}}
     \toprule
     \textbf{Data science step} & 
        \textbf{Definition (taken from~\cite{ramasamy2023workflow})}&
     \textbf{\% of events} \\
      \midrule
      \rowcolor{gray}
     \texttt{data\_preprocessing} & The process of preparing a dataset(s) for the subsequent analysis. It includes tasks such as cleaning, instance selection, normalization, data transformation, and feature selection. & 37.1\%\\
     \texttt{data\_exploration} & The process of inspecting the content and shape of a dataset to understand the nature and characteristics of the data. Note that it may involve the usage of visualization techniques but differs in its purpose. & 29.8\%\\

    \rowcolor{gray}
    \texttt{comment\_only} & Lines of comments including commented code. & 11.2\%\\
          
    \texttt{modelling} & The process of applying statistical models and learning-based algorithms to learn from sample data. & 7.3\%\\
    
    \rowcolor{gray}
    \texttt{helper\_functions} & Code that is not directly related to the data science activity at hand but provides useful scripting functions (\textit{e.g.}, importing or configuring libraries). & 6.3\%\\

    \texttt{load\_data} & The process of loading a dataset of any type (\textit{e.g.}, .csv, .pkl) into a Jupyter notebook environment. & 5.5\%\\

     \rowcolor{gray}
    \texttt{evaluation} & The process of assessing a model using one or more evaluation metrics such as goodness of fit and accuracy. & 1.7\%\\

    \texttt{prediction} & The process of applying a model trained on a set of data to other or newly arriving pieces of data to forecast new values.  & 1.0\%\\

    \rowcolor{gray}
    \texttt{result\_visualization} & The process of obtaining a graphical representation (e.g., tables, plots, graphs) of measurements & 0.1\%\\
    
    \texttt{save\_results} & The process of serializing and storing the data. & 0.1\%\\

    \bottomrule
    \end{tabular}
    \caption{Data science steps and their distribution (in the descending order) in the dataset.}
    \label{tab:data-science-step-distribution}
\end{table}

To understand the evolution of the notebooks in the context of the data science workflow, we annotated our data with labels of various data science steps. To generate the labels, we use the characterization of data science steps in computational notebooks provided by Ramasamy et al.~\cite{ramasamy2023workflow}. A complete list of data science steps, their definitions, and their amount in our data can be found in Table~\ref{tab:data-science-step-distribution}.

To identify the data science step for each logged cell in our data, we developed a single-label classifier based on the DASWOW dataset~\cite{ramasamy2023workflow}. Originally, the DASWOW model was designed to analyze standalone notebooks in a multi-label setup, which made it more challenging to use in our case, as it complicates the analysis of transitions between distinct data science steps. To address this limitation, we made several key architectural modifications. First, we converted the multi-label classification problem into a single-label classification task by training only on the \textit{primary\_label} from DASWOW, ensuring each cell receives exactly one data science step label. Second, we replaced the original feature extraction approach with pre-trained CodeBERT embeddings~\cite{feng2020codebert}, which provide robust semantic representations of code. Third, we employed CatBoost~\cite{prokhorenkova2018catboost} as the classification algorithm instead of the original architecture, which provided better performance and more reliable predictions without requiring a certainty threshold that would leave some cells unlabeled.

We trained the CatBoost classifier with 400 iterations, a learning rate of 0.2, and weighted F1 score as the evaluation metric, similar to the original paper. The model was validated using 5-fold stratified cross-validation on the DASWOW training set, achieving a cross-validated F1 score of 0.71±0.02. On the DASWOW test set, we achieved an F1 score of 0.721, even slightly higher than the original model's F1 score of 0.716. Importantly, unlike the original model, which employed a certainty threshold and could not label all examples, our model can classify every cell with a single label, making it more suitable for transition analysis.

To understand whether the data science steps change at the cell level during development, we studied the probabilities of transitions between various pairs of data science steps. 
To investigate the distribution of data science steps over time, we divided the log data for a given notebook into ten quantiles sorted by time, and then counted the steps for each bin. We hypothesize that certain steps may be more prevalent at the start or the end of the development process in Jupyter notebooks, \textit{e.g.}, that in general data exploration precedes modelling.

In this research question, in addition to analyzing self-transitions, we include an analysis of inter-transitions. Examining the differences in the process between working within a single cell and working across multiple cells can be insightful. We hypothesize that there is minimal switching between data science steps during self-transitions, while inter-transitions will demonstrate more frequent transitions between different steps, \textit{i.e.}, people are less likely to switch their step of the workflow while being in the same cell. The reason for this is that if cells can be interpreted as proto-functions~\cite{titov2022resplit}, then they should be more logically consistent internally than between different cells. In contrast to self-transitions, where we normalize the probabilities for each class, for inter-transitions, we ensure that the sum of all elements in the matrix is equal to $1$. By doing so, we can interpret the probability associated with each transition (for each cell in the matrix) as the likelihood of that specific transition occurring during the development process. This provides insights into the significance of particular transitions within the total number of transitions during development.
\section{Findings}
\label{sec:findings}

\subsection{\textbf{RQ1:} Nature of Code Changes}

\begin{figure*}
    \centering
    \includegraphics[width=\textwidth]{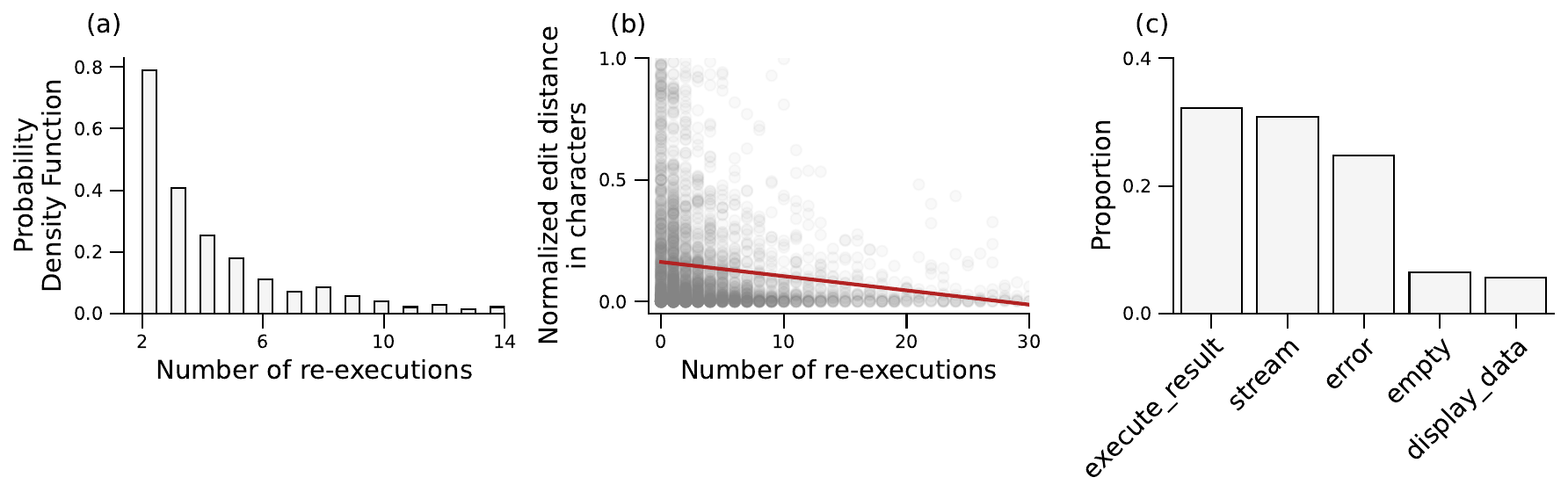}
    \caption{(a) Distribution of the number of consecutive re-executions. Note the exponential nature of the distribution — cells with only one re-execution account for 25.6\% of all re-executions. (b) Correlation between the normalized edit distance between re-executions and the number of consecutive re-executions. Cells with a higher number of re-executions tend to have smaller edit distances, which may indicate a debugging process. (c) Distribution of output types prior to re-execution. Error outputs represent only one third of the outputs in cells before re-execution, suggesting that re-execution is more often related to semantic bugs rather than syntactic ones.}
    \label{fig:reruns}
\end{figure*}

We found that self-transitions account for 39\% of all transitions, while inter-transitions account for the remaining 61\%. As can be seen in the presence of multiple loops in the example evolution graph (see Figure~\ref{fig:example}), self-transitions frequently occur in series. Figure~\ref{fig:reruns}a displays the distribution of the number of subsequent self-transitions, revealing that the mean number of re-executions for cells that have been re-executed at least once is $5$. Additionally, 5\% of the cells with at least one re-execution account for 25.6\% of all re-executions. This shows that a relatively small percentage of code in notebooks demands a significant amount of effort from users, which might indicate the need for special tooling.

To explore the changes further, we plotted the correlation in Figure~\ref{fig:reruns}b between the number of re-executions and the mean normalised edit distance in symbols per re-execution. From this plot, we observe that the number of changes decreases over time (\APAr{0.2}{0.01}), thus indicating that a large number of re-executions typically reflects iterative tweaking of the code, such as debugging or exploration, rather than a development process. The mean change size between two consecutive runs is approximately 13\% of the cell size. Based on this small change size, it seems reasonable to think that this corresponds to a fix or the addition of a single statement.

\sloppy
Lastly, we analyzed the distribution of types of output before self-transitions. Figure~\ref{fig:reruns}c shows: (1) \textit{execute\_result} -- rich output of the cell, \textit{e.g.}, dataframes; (2) \textit{stream} -- text output from print and other stream sources; (3) \textit{error} -- error messages from interpreter; (4) \textit{empty} -- no output for the cell; (5) \textit{display\_dat}a -- visualization from matplotlib and other graphics.

The most common output prior to cell re-execution is rich output, such as dataframes. This is likely because dataframes were central to both tasks, and participants were primarily focused on transforming and validating these objects as part of their workflow. The second most common output was stream outputs, likely driven by exploration, although a significant portion of these texts could be related to debugging. Supporting this hypothesis is that more than 20\% of outputs prior to re-execution are error messages. Lastly, the least common outputs were visualizations and empty outputs. Empty outputs likely represent the development process, where people are making blind code modifications, while visualizations are relatively rare, even in notebooks.

\observation{\textbf{Summary of RQ1}: In case of self-transitions, we find that most changes introduced in iterative work over singular cell are relatively small yet repeated multiple times in a cyclical manner. This may indicate that the interactive nature of notebooks influences the development process towards smaller experimental changes for debugging and exploration.}

\subsection{\textbf{RQ2:} The Purposes of Code Changes}

We analyzed the set of labels provided by GPT-4o for the $3,573$ self-transitions. When we revised the sample of $400$ transitions to come up with a final ``ground truth'' label, we established that GPT-4o showed a $0.55$ overlap with the ground truth over these $400$ transitions, in our 11-label task. We compute this overlap by calculating the number of transitions for which GPT-4o and the ground truth share at least one label in common. However, there are some non-overlapping answers for which the GPT-4o label could also be considered as appropriate. For instance, GPT-4o identified $327$ ``no change''-s, while technically, a string comparison revealed that only $306$ cases were no-code changes. $7$ additional transitions contained changes that could potentially be semantically considered a ``no content change'' (\textit{e.g.}, removing or adding a newline character). The remaining $14$ cases contained changes that can only be considered code changes.

 \begin{figure}[t]
    \centering
    \includegraphics[width=0.8\textwidth]{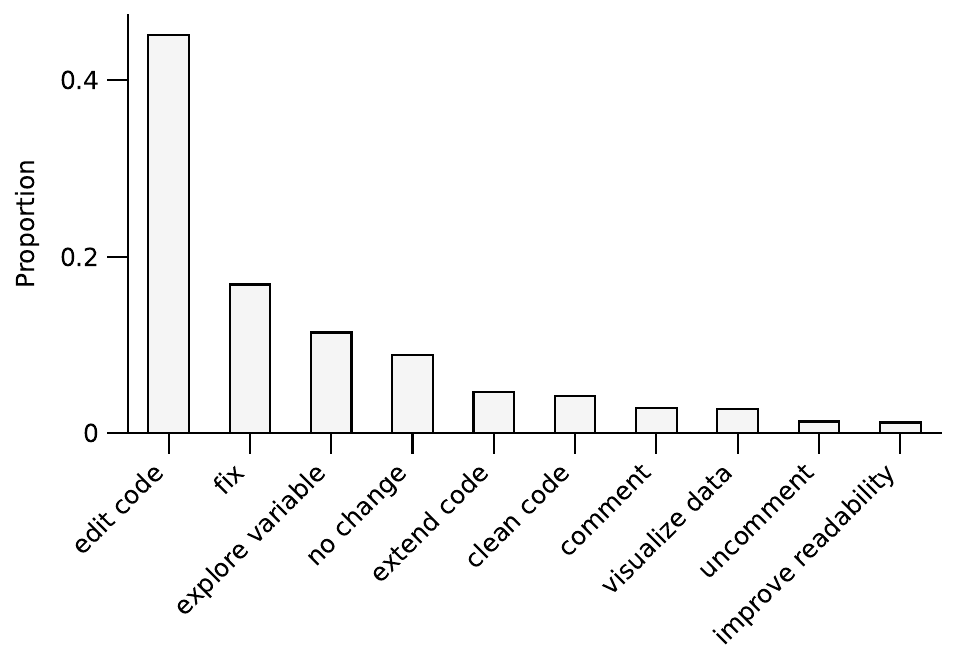}
    \caption{Proportion of labels provided by GPT-4o (as of the 10th of March, 2025) when asked to classify the goal that the developer pursued when implementing the code change(s) in each code transition. Note that some code transitions may include more than one change purpose and labels with frequency less then 1\% were omitted from the plot.}
    \label{fig:barplot_labels_gpt}
\end{figure}

Figure \ref{fig:barplot_labels_gpt} shows the proportion of occurrence of the purpose labels. 
8.8\% of the labels refer to ``no change''. These transitions represent the cells that developers re-executed without editing the content. This could be a sign that the user needed to re-execute a cell due to technical problems, like an unsuccessful rendering of the output. As for the other labels, we can group them into three major categories:
\emph{code iteration} (``fix'', ``debug'', ``edit code'', ``clean code'', ``improve readability'', ``comment'', and ``uncomment''), \emph{exploration} (``explore variable'' and ``visualize data''), and \emph{further development} (``extend code''). Let us now go over them one by one, referring to Figure~\ref{fig:barplot_labels_gpt} in all cases.

\paragraph{Code iteration} This category is present in 72.2\% of the labels. ``Fix'' (fixing syntactic errors) alone represents 16.8\% of the label count. This means that developers needed to re-execute cells to identify the syntactic mistakes they made while writing their code. Such a result suggests that the support that these developers received in Jupyter might not be sufficient or effective in terms of Python/library code documentation and syntax control. 
The more lines and the more complexity the cells contain, the more inefficient this way of discovering errors becomes. This issue could be addressed by extending Jupyter with an assistant that helps developers auto-complete code and check syntax correctness on the fly, similarly to other IDEs.
The ``edit code'' label also appeared with high frequency --- in 45.1\% of labels, pointing to the need that developers seem to have to iterate and gradually implement, as they need the execution feedback. 
The results also show that developers do ``clean code'' and change code to ``improve readability'', which indicates that they care about the final code quality. Given the competitive context of the hackathon, it makes sense to find instances of these labels. 

\paragraph{Exploration} This category appears in 14.1\% of the labels. 
The label ``explore variable'' occurred in 11.4\% of the labels, while ``visualize data'' --- in 2.7.\%. This indicates that the features for inspecting variable content in Jupyter are not sufficient.
 
\paragraph{Further development} This category, including the single label ``extend code'', is present in 4.7\% of the cases. The fact that developers need to re-execute cells after adding new pieces of code shows that they need to gradually verify the performance of their code. This also calls for more IDE features in Jupyter notebooks.

\observation{\textbf{Summary of RQ2}: Based on the provided annotations, it is visible that the interactive nature of notebooks is often leveraged for code iteration purposes, including tasks for editing and fixing code. This result suggests a lack of adequate tools to support the code iteration in the notebook environment.}

\subsection{\textbf{RQ3:} Changes in Data Science Steps}

We first analyze when each data science step occurs, and then we study the transitions between data science steps during the development. 

\subsubsection{Stages of different data science steps} 

\begin{figure*}
    \centering
    \includegraphics[width=\textwidth]{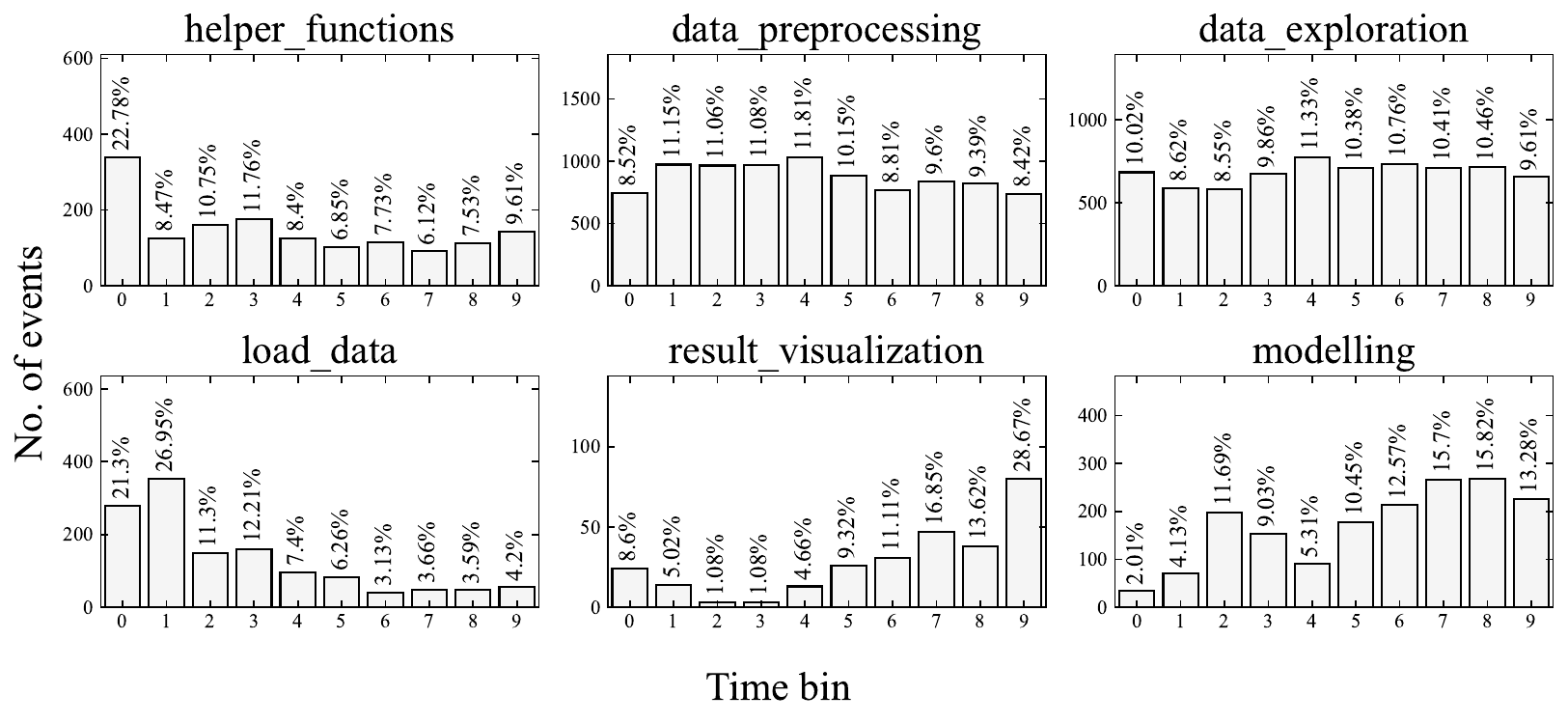}
    \caption{The presence of different data science steps depending on the quantile of the history of a notebook.}
    \label{fig:dsstep_quartiles}
\end{figure*}

With the help of the generated data science labels, we investigate when each of the data science steps occurs in a notebook development process.
We categorize the log data for a given notebook into ten quantiles sorted by time. In Figure~\ref{fig:dsstep_quartiles}, we plot the distribution of labels across the quantiles. Our results show that while some steps occur as expected (\textit{e.g.}, \textit{load\_data} events occur more frequently in the first few quantiles, and \textit{modelling} events occur more frequently in the last few quantiles), there are some surprising trends. We find that the \textit{helper\_functions} step, which is about importing libraries and defining custom functions, occurs fairly evenly throughout the development process, instead of happening at the start as one might assume. Similarly, \textit{data\_preprocessing} and \textit{data\_exploration} steps occur throughout the entire development process and not only towards the first half. 

This provides additional support for the findings of Chen et al.~\cite{chen2025towards} that the \textit{data\_preprocessing} and \textit{data\_exploration} steps are the most error-prone, since they constitute the majority of actions in realistic Jupyter Notebook development. Interestingly, their third most frequent error-related step, \textit{load\_data}, is rare in our data, suggesting that it is relatively difficult for users.

These results demonstrate that the process of development could be split into time-dependent steps like \textit{load\_data} or \textit{modeling} and time-independent steps like \textit{data\_preprocessing} and \textit{data\_exploration}. This classification can aid in developing tools for notebooks. When constructing tools to support data exploration or Extract Method refactorings (to extract helper functions), it would be beneficial to design these tools for continuous assistance, operating after each action in the notebooks. On the other hand, tools assisting with modeling or loading data could be invoked only several times when the corresponding action is detected. 

\subsubsection{Transitions between data science steps} 

To answer this question, we look at the transitions of data science steps between notebook cells. Based on the algorithm described in Section~\ref{sec:method:workflow}, we generate a transition matrix of probabilities for data science steps in a workflow. As stated in the methodology, we focus on both self-transitions and inter-transitions.

\paragraph{Self-transitions} 

The number of self-transitions in our data accounts for 39\% of all executions. This confirms the cyclical nature of development in Jupyter notebooks, as users continuously iterate over the same cell and experiment with the code contained within it. 
We arranged self-transitions into transition probability matrices.
The notable characteristic of the resulting matrices is their diagonal dominance, which indicates that---in the vast majority of instances---the data science label remains unchanged after re-executing the cell (\textit{i.e.}, the step does not change across re-executions). 
This feature underscores both the stability of our machine learning annotation model and the infrequent need for users to carry out multiple stages in the same cell. 

\begin{figure}
    \centering
    \includegraphics[width=0.95\textwidth]{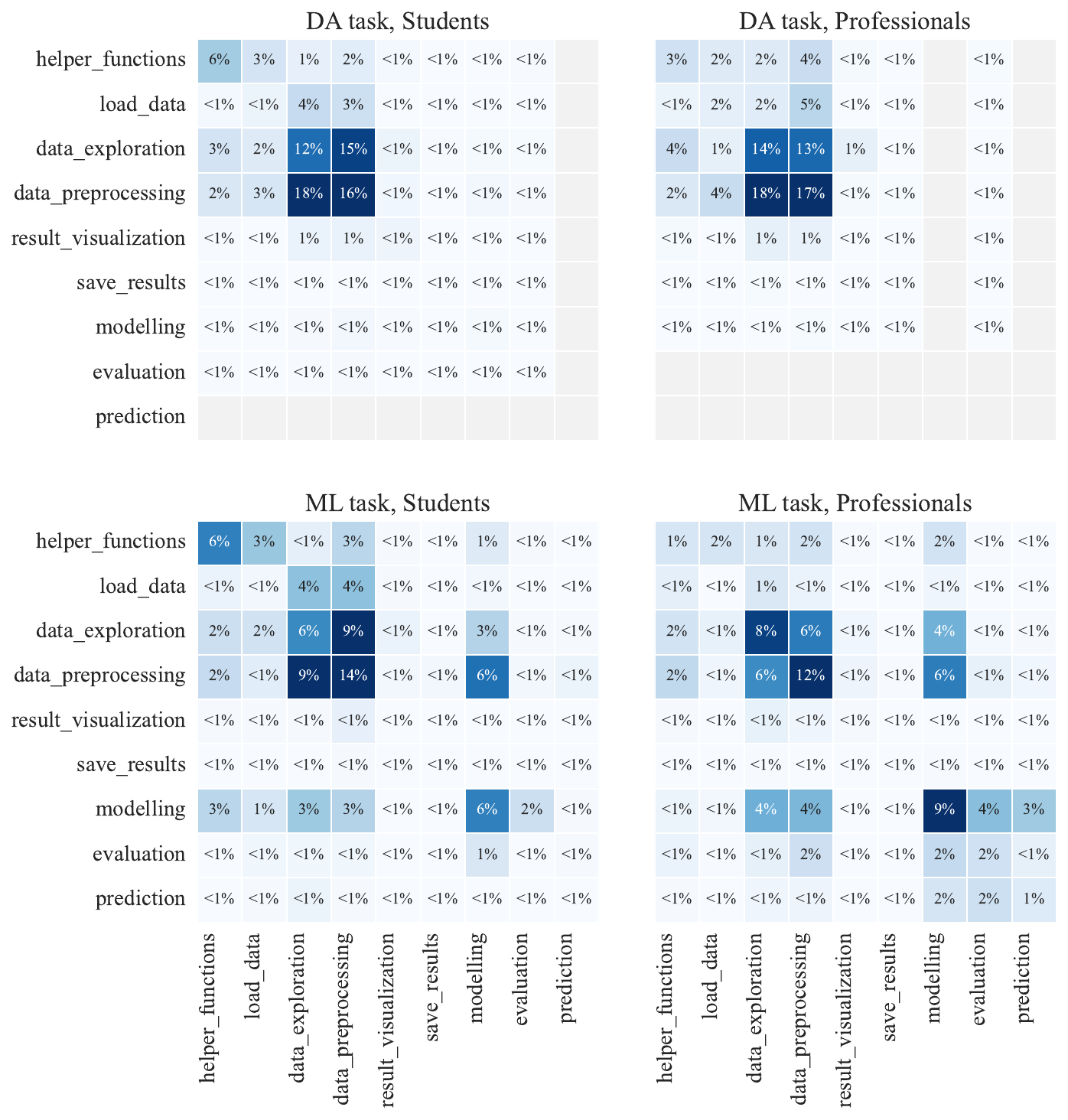}
    \caption{The transitional matrices for data science steps --- only inter-transitions. Note that the matrices are normalized so that all the probabilities sum up to 1.}
    \label{fig:transitions}
\end{figure}

\paragraph{Inter-transitions} We continue our analysis by examining transitions between different cells. 
Figure~\ref{fig:transitions} shows the transition matrices for different cells, normalized to sum up to 1. From the matrices, it is apparent that professionals transitioned more from data exploration to helper functions in the DA task. This could be an indication that professionals try to extract methods for data exploration into helper functions. 
For both tasks, a substantial part of transitions are tied between \textit{data\_exploration} and \textit{data\_preprocessing} labels (including transitions to cells with the same label) --- 61\% for the DA task and 35\% for the ML task. This suggests that users engage in data exploration and data cleaning at different stages of the development process. We believe the transition analysis indicates that the main work in the notebook, regardless of the task, revolves around data manipulation. 

\observation{\textbf{Summary of RQ3}: The analysis of the transition matrices between various data science steps indicates that, regardless of the task or level of expertise, core data science steps in the notebooks predominantly involve data exploration and pre-processing.
Some steps are time-dependent, while others are time-independent. Data exploration and processing occur uniformly throughout the entire working period within the notebook. Conversely, other steps tend to occur more frequently towards the start or the end of the notebook development process.}

\section{Discussion}
\label{sec:discussion}

\subsection{Threats to Validity}
\label{sec:ttv}

The main potential challenge to the validity of our conclusions is the reliability of the annotation models. We employed our own model for the data science step annotation, and while it performs slightly better than the one proposed in the original paper, the F1 score is still not very high. The annotations for the change types were obtained using GPT-4o. Although the model is incredibly powerful, we cannot fully trust the results produced by this annotation. However, the conclusions drawn from this analysis are based on strong statistical effects, so we believe that the quality of the model does not significantly impact the results.

It is also important to note the sample size in our study. Although our data collection led to a considerable number of actions within notebooks, it was based on a relatively small sample size of 20 developers, due to the challenge of recruiting participants for user studies, especially those that strive to be realistic and thus take many hours. While we made our best effort to balance the sample in terms of experience, one must be cautious when generalizing our results --- they could be skewed. 

Finally, this project involves several software components: a JavaScript plugin, a Python server, machine learning annotation, and complex data analysis. 
Because the solution is composed of multiple components written in different languages, the application may not be stable on newer versions of Jupyter Notebooks. We have released all of our code, datasets, and materials to the community so they can be used for future research, and we also provide pre-calculated results and intermediate steps for the supplied notebooks so the results can be reproduced and verified step by step~\cite{artifacts}.

Despite all these important limitations that grow from the novelty of our work, we believe that they do not invalidate the general high-level observations of our study.

\subsection{Implications}

From the results of the study, we can summarize the following implications and action points for notebook tool developers and researchers. 

\textbf{Jupyter Notebook developers need more code iteration tools.}
A significant amount of cyclical work in notebooks is related to code iteration, as users tend to iteratively edit and fix their code, leveraging the interactive nature of the notebooks and data-driven tasks. To assist with this and bring notebooks closer to their original goal of being a tool for exploration, we suggest adding more debugging tools. Features like IDE-style inspections, variable tracking, or even dedicated debugging cells that are excluded from the linear structure of the notebook could greatly enhance the user experience. There are already several Jupyter Notebook tools that support extended debugging experience~\cite{patra2022nalin,grotov2024debug,yang2022data}. However, like many Jupyter plugins, they suffer from a lack of popularity and therefore limited adoption. We hope that our study will help highlight the importance of such features, and that some of them may eventually be supported by the Jupyter community or by major IDE providers such as Microsoft or JetBrains.

\textbf{The data science workflow is highly entangled}. Our analysis shows that much of the workflow involves frequent transitions between exploration, preprocessing, and visualization. We argue that the current  classification model does not accurately capture this process, as many steps occur only a few times or not at all, while the remaining steps are closely intertwined. Additionally, a significant amount of work happens within the development of individual cells, which is not accounted for in the existing classification. To better understand the data science process, a more refined taxonomy is needed. The next iteration of the taxonomy and classification model should also account for developer-related variables, such as experience or background. Since we collected data from both professionals and students and labeled it accordingly, our dataset may provide a foundation for this line of work.

\textbf{Snapshots of notebooks are not representative of the actual development process.} Our work demonstrates that a significant portion of the development occurs within individual cells. This development cannot be fully analyzed using VCS snapshots of the notebook, which could miss many important insights about the process. We show that users interact with the code in various ways during cell development, and a deeper analysis of these interactions could lead to much better support for notebook development. 
Following the work of Chen et al.~\cite{chen2025towards}, we are extending the corpus of fine-grained Jupyter log data. The combined datasets may not only improve our understanding of the Jupyter Notebook development process, but also create new opportunities in the field of machine learning. More available data could enable more precise models of data science workflows and support models for next-edit suggestion. We hope that our tooling is used to collect even more datasets in the future.

\section{Possible Future Research}
\label{sec:fututre}

In this section, we showcase several directions where one can go to investigate our data further. We also highlight some relevant results that did not fit our main research direction but that can instead be used as a foundation for future deeper exploration.

\subsection{Temporal Analysis of the Workflow} 

One of the strengths of the dataset we collected is that it contains timestamps of each action. It could be interesting to take a look at how people spent their time during the development process. For example, we can calculate the amount of time spent on the execution state of cells, which accounts for an average of 8.96\% of the total time spent in the notebook. Table \ref{tab:time_stats} presents a summary of the time users spend in the execution state for each task, along with its percentage in relation to the overall task solving.

\begin{table}[t]
    \centering
        \begin{tabular}{llcc|cc}
        \toprule
         \multirow[h]{2}{*}{\textbf{Task}} & \multirow[h]{2}{*}{\textbf{Expertise}} & \multicolumn{2}{r}{\textbf{Execution time (s)}} & \multicolumn{2}{r}{\textbf{\% of total time (\%)}} \\\cmidrule(lr){3-4}\cmidrule(lr){5-6}
         &  & \textbf{mean} & \textbf{std} & \textbf{mean} & \textbf{std} \\
        \midrule
        \multirow[h]{3}{*}{ML} & Student & 7.63 & 50.74 & 11.90 & 9.50 \\
         & Professional & 5.47 & 46.53 & 14.21 & 13.35 \\
         & All & 6.33 & 48.26 & 13.05 & 11.12 \\
        \midrule
        \multirow[h]{3}{*}{DA} & Student  & 2.95 & 15.81 & 7.60 & 8.48 \\
         & Professional  & 1.22 & 9.20 & 2.94 & 2.58 \\
         & All & 2.58 & 14.67 & 6.38 & 7.61 \\
        \midrule
        \multirow[h]{3}{*}{All} & Student & 4.05 & 28.26 & 8.89 & 8.78 \\
         & Professional & 3.90 & 37.39 & 9.08 & 11.24 \\
         & All & 3.99 & 31.84 & 8.96 & 9.54 \\
        \bottomrule
\end{tabular}
    \caption{Descriptive statistics for the execution time and the percentage of execution time to total time.}
    \label{tab:time_stats}
\end{table}

Looking at the data, we can observe a difference in the average time taken for cell execution between the DA task \APAstats{2.58}{14.68} and the ML task \APAstats{6.33}{48.26}, as well as between students \APAstats{4.05}{28.26} and professionals \APAstats{3.90}{37.39}. This is an interesting finding that shows a difference in the way students and experts engage in various data science tasks. However, to understand the reasons that lead to such a difference would require a much deeper analysis. One could analyze the dataset further to identify the actions that take most of the time to execute, how the time between actions is distributed, or even predict the execution time from the cell source. All the details for the temporal analysis and several other features for time calculation can be found in supplementary materials~\cite{artifacts}. 

\subsection{Analysis of Code Dynamics} 

\begin{figure}[ht]
    \centering
    \includegraphics[width=\textwidth]{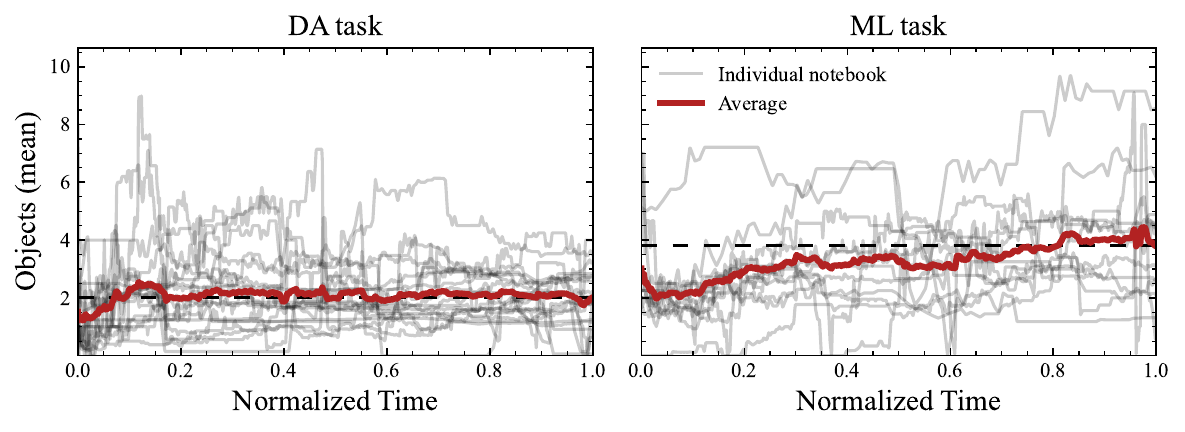}
    \caption{The evolution of the \textit{Mean number of objects} metric in time for individual notebooks (grey lines) and their average curve (red line).}
    \label{fig:objects_evolution}
\end{figure}

Another interesting topic to investigate further is how the code evolves during the development of the notebook. One could extend our analysis by looking at, for instance, how code complexity varies over time or identify the most frequent errors during the development of these notebooks. 
We calculated how the mean number of Python objects (nodes of the abstract syntax tree with a name and a value) per cell changed during the development of the notebook (see Figure~\ref{fig:objects_evolution}). We can clearly see the difference in dynamics between the two tasks. We believe that there should be other interesting effects that require deeper investigation. One can pursue studying what errors frequently occur in the notebooks, what objects are most frequently changed, and how exactly people change code when they re-run multiple cells in a row. All of the details for code analysis and several other code-based features can be found in supplementary materials~\cite{artifacts}. 

\section{Conclusion}
\label{sec:conclusion}

This study provides a unique perspective on the real-time operation of Jupyter notebooks. We demonstrate that analyzing fine-grained logs in notebooks can provide valuable insights for tool developers and researchers. By focusing on code changes, we reveal that a significant portion of user effort is directed towards debugging rather than exploration. This calls for work on improved debugging tools in Jupyter notebooks.

We further analyzed the development in terms of the framework modelling data science steps proposed by Ramasamy et al.~\cite{ramasamy2023workflow}, finding that the dynamics of real-time data science workflow in our study align with the proposed steps, but without necessarily following a linear structure.

Consequently, we propose that the development of tools and notebook features should prioritize supporting the process over merely focusing on the final form of the notebook. Embracing and facilitating the inherent non-linearity of the development process, rather than attempting to counteract it, should be a key objective in future tools for Jupyter notebooks.

\section*{Data Availability}

Our materials, code, the dataset, and the extended results can be found in our supplementary materials~\cite{artifacts}. There, one can also find all the necessary tools to conduct further studies on the presented datasets, including those proposed in Section~\ref{sec:fututre}, as well as tools for collecting dynamic Jupyter notebook data and performing similar analyses. Additionally, the repository contains the code and intermediate results required to fully replicate the findings of the presented study.
\section*{Acknowledgments}

This work was partially supported by the Swiss National Science Foundation through projects `D3' (contract no.\ CRSII5\_205975) and `CrowdAlytics' (contract no.\ 200020\_184994). A. Bacchelli gratefully acknowledges the support of the Swiss National Science Foundation through the SNSF Project 200021M\_205146.
We would also like to thank the hackathon participants for their valuable contribution.

\bibliographystyle{elsarticle-num}
\bibliography{paper}
\end{document}